\documentstyle[preprint,aps]{revtex}
\begin{document}
\draft
\title {\bf Two-loop QED with External Magnetic Field }
\author{Pradip K Sahu}
\address{Theory Group, Physical Research Laboratory, Ahmedabad
380 009,
India;\\
E-mail: pradip@prl.ernet.in.}
\maketitle
\begin{abstract}
In two - loop effective Lagrangian, the low - temperature expansion
of the $QED_{3+1}$ with a constant magnetic field and a finite
chemical potential is performed. We then calculate the total
fermion density, some components of polarization operator and de
Hass -van Alphen oscillations. We find that there is a
significant contribution from two-loop expansion to magnetization
and fermion density for higher values of chemical potentials.
\vskip 0.2in
\end{abstract}
\section{Introduction}
In the last two decades, it has been argued that in the stellar objects
such as surface of neutron stars\cite{shap,chan},
supernovae\cite{ginz,mull} and white magnetic
dwarfs\cite{chan,ange}, the magnetic field
strength and the fermion density is very high. Therefore,
the corresponding quantum corrections are important\cite{shab} to such
astrophysical objects. From the
sample of more than 400 pulsars, the range of the
surface  magnetic field strength is in the range of $2\times 10^{10}$ G
$\le H\le 2\times 10^{13}$ G\cite{manc}. Though the magnetic fields
permeate most astrophysical systems, but their origin is
not known. There are two different physical mechanism leading to an
amplification of some initial magnetic field in a collapsing stars.
The one proposed by Bisnovatyi-Kogan\cite{bisn} is due to differential
rotation and other one is due to a dynamo mechanism proposed by Thompson
and Duncan\cite{dunc}.
The magnetic fields as strong as $H
\sim 10^{14}~~- 10^{16}$ G, or even more, might be generated
in new born neutron stars. Recently, it has been argued that
at the time of the primodial nucleosynthesis for the primodial
Big-Bang plasma, the amplitude of magnetic field fluctuations
can be as large as $H$$\simeq 10^{14}$ G\cite{taji}.
A magnetic fields upto the order $H$$\simeq 10^{17}$ G\cite{nara}
also has been suggested for extra galactic gamma bursts in
terms of mergers of massive binary stars.
Even larger macroscopic magnetic fields ($H \ge
10^{18}$ G) can be contemplated are superconducting strings\cite{bere},
which are more speculative systems.
Very recently\cite{vach}, it has been suggested that if the Higgs field
possesses different electro-weak phases in neighboring regions
may produce a magnetic field ($H \sim 10^{21}$ G).
Since, the plasma of thermal equilibrium can sustain fluctuations
of the electromagnetic fields, it is advisable to study
the $QED$ in such high magnetic fields.
\par
In the present paper, we consider the constant and uniform
magnetic fields ($H$) within $QED$ with chemical potential $\mu$,
temperature $T$. The one-loop effective $QED$ Lagrangian ${\it
L}^{eff}(H,\mu,T)$ in
the finite temperature and density is intensively
discussed\cite{elmf}.
Also, it has been discussed and derived to obtain a
low-temperature expansion of the one-loop effective Lagrangian
for a wider range of parameters $\mu$ and $H$\cite{zeit}. The fermion
density, the magnetization, the Hall conductivity and some
components of the polarization operator in the static limit
$p_o=0$, $\vec p\rightarrow 0$ have also been derived\cite{zeit}.
Performing the expression upto two-loop effective
Lagrangian, we shall obtain temperature corrections to the
fermion density and de Haas-van Alphen oscillations for $T^2\ll eH$,
$T^2\ll \mu^2 -m^2$.
Then following the Ref.\cite{zeit}, we shall
calculate some components upto the two-loop polarization operator
in the static limit as well as Hall conductivity in the $QED$.
Here, we have considered only the temperature shifts in the energy of an
electron or positron caused by the interaction with the plasma
electrons and the plasma positrons in two -loop corrections. In the
two -loop expressions fo simplicity, we neglect the radiative
shift of the energy of an electron or a positron in a constant
magnetic field and shift in the energy of an electron or
a positron caused by the interaction with equilibrium radiations.
\section{Low -temperature corrections}
The $QED$ Lagrangian with uniform magnetic field $H$ at
finite density, and at finite chemical potential $\mu$ is:
\begin{equation}
{\cal L} = -\frac{1}{4} F_{\mu\nu}F^{\mu\nu} +\bar{\psi}(i {\partial
\kern-0.5em/} -e {A\kern-0.5em/}  - \gamma_0 \mu -m)\psi~~~.
\label{eq1}
\end{equation}
\noindent
Here, we choose the  external magnetic field to be  parallel to
the $z$--axis, $F_{12}=-F_{21}=H$.

The effective Lagrangian at temperature
$T,~\mu,~H \ne 0$, ~${\cal L}^{{\rm eff}} (H, \mu, T)$
is written as\cite{elmf,zeit}:

\begin{equation}
 {\cal L}^{{\rm eff}} (H, \mu, T) =   {\cal L}^{\rm eff}(H)+
 \tilde{\cal L}^{\rm eff}(H,\mu,T),
\label{eff}
\end{equation}

\noindent
where $\tilde{\cal L}^{\rm eff}(H,\mu,T)= {{\cal L}^1}^{{\rm eff}}(H, \mu, T)
 +   {{\cal L}^2}^{\rm eff}(H, \mu, T)$, correspond to one-loop
and two-loop effective Lagrangian.
The one -loop effective Lagrangian is given in Ref.\cite{zeit}, which is:
    \begin{equation}
        {{\cal L}^1}^{\rm eff}(H,\mu,T)=
        \frac{eH}{2\pi^2} \sum_{n=0}^\infty
        b_n \int_{-\infty}^\infty d\! p_{z}
{}~\frac{p_{z}^2}{\varepsilon_n(p_{z})}
        ~(f_+(T) + f_-(T)).
        \label{eql1}
        \end{equation}
\noindent
Whereas, we follow the procedure as given in Ref.\cite{zhuk} to derive
the two - loop effective Lagrangian, which may be written as:
    \begin{equation}
        {{\cal L}^2}^{\rm eff}(H,\mu,T)=
        \frac{eH}{2\pi^2} \frac{\alpha}{2\pi} \sum_{n=0}^\infty
        b_n \int_{-\infty}^\infty d\! p_{z}~p_{z} [\ln (\frac{p_{z}^2}{eH})+1]
        ~(f_+(T) + f_-(T)).
        \label{eql2}
        \end{equation}
\noindent
These expressions are  the contribution due to the finite
temperature and density. Here, $p_{z}$ is the momentum
parallel to the magnetic field and  $b_n\equiv 2- \delta_{n0}~$,
for $n=0$, $b_0=1$ is the lowest Landau level.
In Eq. (\ref{eff}),  ${\cal L}^{\rm eff}(H)$ is the Schwinger  Lagrangian
in  the purely magnetic case\cite{schw}:

        \begin{equation}
        {\cal L}^{{\rm eff}} (H)=-\frac{1}{8\pi^2} \int_0^\infty \frac{ds}{s^3}
        \left[
eHs \coth(eHs)-1-\frac13 (eHs)^2
\right]
        \exp(-m^2s).
\label{eqsc}
        \end{equation}

\noindent
In the expressions Eq.(\ref{eql1}) and Eq.(\ref{eql2}), $f_\pm(T)$ represents
 the Fermi distribution,

        \begin{equation}
        f_\pm(T) = \frac1{1+e^{\beta(\varepsilon\mp \mu)}} \quad .
        \nonumber
        \end{equation}
Since we are considering the low-temperature limit of the $QED_{3+1}$,
in the zero temperature limit ($T \rightarrow 0$), we can replace the
Fermi distribution to the step-function, $\lim_{T \rightarrow 0} f_\pm
= \theta(\pm\mu - \varepsilon)$ and Eq. (\ref{eql1},\ref{eql2}) may be
written as:

        \begin{eqnarray}
        \lefteqn{{{\cal L}^1}^{\rm eff}(T = 0,H,\mu)=}
        &&\label{eqT0}\nonumber \\
        &&  \frac{eH}{2\pi^2}
        \sum_{n=0}^{ \left[   \frac{\mu^2 - m^2}{2eH}   \right]}
        b_n
        \left\{
\mu  \sqrt{\mu^2-m^2-2eHn}
        - (m^2+2eHn)
        \ln
        \frac{ \mu + \sqrt{\mu^2-m^2-2eHn}} {\sqrt{m^2+2eHn}}
\right\},
\nonumber\\
\nonumber \\
        \lefteqn{{{\cal L}^2}^{\rm eff}(T = 0,H,\mu)=}
        && \\
&&
  \frac{eH}{2\pi^2}\frac{\alpha}{2\pi}
        \sum_{n=0}^{ \left[   \frac{\mu^2 - m^2}{2eH}   \right]}
        b_n
        \left\{
(\mu^2-m^2-2eHn) \ln
        \frac{ \mu^2-m^2-2eHn} {eH}
\right\}.
\nonumber
        \end{eqnarray}

 \noindent
Where the upper limit of $n$ sum may be obtained from the relation
$\mu^2 - m^2 -2eHn\ge 0$.
\par
To evaluate the low-temperature corrections to the effective
Lagrangian (\ref{eql1}, \ref{eql2}), we follow the procedure
as given in Ref.\cite{zeit}. First we take the derivative of
$\tilde{\cal L}^{\rm eff}(H,\mu,T)$ with respect to $T$, with
a fixed chemical potential. After some purely algebraic manipulations
and integrating with respect to temperature, we obtain the
temperature corrections to the zero temperature Lagrangian Eq. (\ref{eqT0}),

\begin{eqnarray}
       \Delta{{\cal L}^1}^{\rm eff}(T,H,\mu)= \frac{eH}{2}\frac{T^2}{6}
        \sum_{n=0}^{ \left[   \frac{\mu^2 - m^2}{2eH}   \right]}
        b_n \frac{\mu}{(\mu^2 - m^2 - 2eHn)^{1/2}}
        + O(T^4),
\nonumber \\
\nonumber \\
        \Delta{{\cal L}^2}^{\rm eff}(T,H,\mu)=
\frac{eH}{2}\frac{\alpha}{2\pi}\frac{T^2}{6}
        \sum_{n=0}^{ \left[   \frac{\mu^2 - m^2}{2eH}   \right]}
        b_n
\left[
1 + \frac{2\mu^2}{\mu^2 - m^2 - 2eHn} + \ln
\frac{\mu^2 -m^2 -2eHn}{eH}
\right]
        + O(T^4).
\nonumber \\
        \end{eqnarray}

\noindent
In the above expressions, we differ a factor of $1/2$ from Ref.\cite{zeit}
for the one -loop corrections.
The above expansions are valid for $ \frac{T}{\mu - \sqrt{m^2 + 2eHn}} \ll 1$.
\par
In figure 1, we present the ratio of total effective Lagrangian
(including two - loop corrections) to one - loop effective Lagrangian as a
function of magnetic field for fixed chemical potential. Where case
(a), (b) and (c) are  for $\mu=$ 2, 5 and 10 MeV respectively.
We noticed that the ratio is high for large value of chemical
potential and it decreases with increase in magnetic fields. The
ratios are quite different from each other for various values of
chemical potentials with variation of magnetic fields.
\par
Now, we can write a low-temperature expansions to the total
effective Lagrangian including both the
one-loop  and two-loop effective Lagrangian:

\begin{eqnarray}
        {\tilde{\cal L}^{\rm eff}(T,H,\mu)}=\nonumber\\
        &&  \frac{eH}{2\pi^2}
        \sum_{n=0}^{ \left[   \frac{\mu^2 - m^2}{2eH}   \right]}
        b_n
\left\{
\mu  \sqrt{\mu^2-m^2-2eHn}
        - (m^2+2eHn)
        \ln
        \frac{ \mu + \sqrt{\mu^2-m^2-2eHn}} {\sqrt{m^2+2eHn}}
\right.\nonumber\\
        & &\nonumber\\
        && \left.
+ \left[
        \frac{T^2\pi^2}{6} \frac{\mu}{(\mu^2 - m^2 - 2eHn)^{1/2}}
        \right]
+\frac{\alpha}{2\pi}
\left[
(\mu^2 -m^2 -2eHn)
\ln \frac{\mu^2 -m^2 -2eHn}{eH}
\right. \right. \nonumber\\
& &\nonumber\\
&& \left. \left.
    + \frac{T^2\pi^2}{6}
\left[
1 + \frac{2\mu^2}{\mu^2 - m^2
- 2eHn} + \ln \frac{\mu^2 -m^2 -2eHn}{eH}
\right]
\right]
\right\}
+ O(T^4).
        \label{total12}
\end{eqnarray}
\section{Fermion density and magnetization}
Using the total effective Lagrangian (\ref{total12}),
one would calculate the fermion density
$\rho = \displaystyle\frac{\partial {\cal L}^{\rm eff}}{\partial
\mu}$, the magnetization
$M=\displaystyle\frac{\partial {\cal L}^{\rm eff}}{\partial H}$,
the Hall conductivity and some components of the polarization
operator in the static limit ~$p_0=0, ~{\bf p} \rightarrow 0$.

Thus the total fermion density is:
$\rho = \displaystyle\frac{\partial {\cal L}^{\rm eff}}{\partial \mu}$ one has:

        \begin{eqnarray}
        \lefteqn{\rho(H,\mu,T) =}\nonumber\\
&& \frac{eH}{\pi^2} \sum_{n=0}^{\left[  \frac{\mu^2 - m^2}{2eH} \right]}
b_n \sqrt{\mu^2-m^2-2eHn}
\left\{
1 - \frac{T^2\pi^2}{12} \frac{m^2 +
        2eHn}{(\mu^2-m^2-2eHn)^2}
\right.\nonumber\\
& &\nonumber\\
&& \left.
+\frac{\alpha}{2\pi}
\left[
\frac{\mu}{\sqrt{\mu^2 -m^2
-2eHn}}
\left(
1 + \ln \frac{\mu^2 -m^2 -2eHn}{eH}
\right)
\right. \right.\nonumber\\
& &\nonumber\\
&& \left. \left.
-\frac{T^2\pi^2}{6}
\mu
\left(
\frac{3(m^2+2eHn)-\mu^2}{{(\mu^2-m^2-2eHn)}^{5/2}}
\right)
\right]
\right\}
    + O(T^4).
\label{rho}
       \end{eqnarray}
\par
In Fig. 2, we show the fermion density as a function of the
chemical potential for fixed magnetic field ($H\approx 10^{15}$
G) with (a) one - loop corrections and (b) one -loop and two
-loop corrections. We noticed that the fermion density is showing an
oscillating behavior as consecutive Landau levels are passing
the Fermi level for all the cases. There is not much
contribution from two -loop corrections to the fermion density
in this case.
\par
The fermion density as a function of magnetic field for fixed chemical
potential ($\mu=10$ MeV) is shown in figure 3. The fermion density is
showing an oscillating behavior as consecutive Landau levels
are passing the Fermi level. The fermion density is higher in case (b)
due to the contribution of two - loop corrections. If one
decreases the value of chemical potential, the two - loop
corrections to fermion density reduces to one - loop corrections, case (a).
In case of neutron stars, we would like to mention that the
chemical potential of fermions are very high\cite{elmf} and hence
the two - loop corrections is significant and important.
\par
The magnetization $M$ is derived from the total effective Lagrangian
(\ref{total12}), which is:

\begin{eqnarray}
        \lefteqn{{M}(H,\mu,T)=}\nonumber\\
        &&
        \frac{e}{2\pi^2}
        \sum_{n=0}^{\left[  \frac{\mu^2 - m^2}{2eH} \right]}
        b_n
        \left\{
        \mu    \sqrt{\mu^2-m^2-2eHn}
         - (m^2+4eHn) \ln
         \frac{ \mu +
        \sqrt{\mu^2-m^2-2eHn}}
        {\sqrt{m^2+2eHn}}
                         \right.\nonumber\\
        &&\nonumber\\
        &&
        \left.
        + \frac{\pi^2T^2}{6} \frac{\mu (\mu^2 - m^2 -
        eHn)}{(\mu^2 - m^2 - 2eHn)^{3/2}}
    +\frac{\alpha}{2\pi}
\left[
\left(
(\mu^2 -m^2 -4eHn) \ln
\frac{\mu^2 -m^2 -2eHn}{eH} -(\mu^2 -m^2)
\right)
\right.\right.
\nonumber \\
&&\nonumber\\
&&
\left. \left.
+ \frac{T^2\pi^2}{6}
\left(
\frac{ 2\mu^2 (\mu^2 -m^2) -2eHn( \mu^2 -m^2 -2eHn)} { (\mu^2
-m^2 -2eHn)^2} + \ln \frac{\mu^2 -m^2 -2eHn}{ eH}
\right)
\right]
\right\}.
        \end{eqnarray}
\noindent
Since, our main aim is to focus the effect of two -loop
contribution, for simplicity, we have neglected the vacuum
magnetization in our calculations.
The magnetic susceptibility $\chi$ is defined as $(\partial
M/\partial H)$. The magnetization and magnetic susceptibility
are useful to calculate the spatial components of polarization
operators\cite{dani}.
\par
Figure 4 shows the magnetization as a function of magnetic field
for fixed chemical potential ($\mu =10$ MeV). We have not
considered the vacuum magnetization in this figure, because the
vacuum contribution is small\cite{elmf}. We noticed that the
fermion gas exhibits the de Hass - van Alphen effect without
and with inclusion of two -loop corrections curve (a) and
curve (b) respectively, which
is in agreement with Ref.\cite{elmf}. The magnetization is high
at small values of magnetic fields and reaches to one - loop
magnetization for large values of magnetic fields. So, the two
-loop contribution to magnetization is significant at particular range of
magnetic fields.
\par
Next, we calculate the some components of the polarization
operator in the static limit. In Ref.\cite{zeit}, it has been
shown that the $\Pi_{00}$ --component of the polarization
operator is nothing but the derivative of the total fermion density
with respect to the chemical potential in the static limit,
e.g., ~$\Pi_{00}(p_0=0,~{\bf p}\rightarrow 0) =
e^2 \displaystyle\frac{\partial \rho}{\partial \mu}$,

        \begin{eqnarray}
        \lefteqn{\Pi_{00}(p_0=0,~{\bf p}\rightarrow
0)=}\label{pi00} \nonumber \\
        && e^2
        \frac{eH}{\pi^2}
        \sum_{n=0}^{\left[  \frac{\mu^2 - m^2}{2eH} \right]}
        b_n
        \left\{
\mu (\mu^2-m^2-2eHn)^{- 1/2}
        + \frac{T^2\pi^2}{4}
        \frac{\mu (m^2 + 2eHn)}{(\mu^2-m^2-2eHn)^{5/2}}
\right.\nonumber\\
& &\nonumber\\
&& \left.
+\frac{\alpha}{2\pi}
\left[
\left(
1 + \frac{2\mu^2}{\mu^2-m^2-2eHn} +
\ln \frac{\mu^2-m^2-2eHn}{eH}
\right)
\right.\right.\nonumber\\
& &\nonumber\\
&& \left.\left.
+ \frac{T^2\pi^2}{6}
\left[
\frac{3(\mu^2-m^2-2eHn)^2-4\mu^2(\mu^2-3m^2-6eHn) }
{(\mu^2-m^2-2eHn)^3}
\right]
\right]
\right\}.
        \end{eqnarray}

In the above expression, we have included the two -loop corrections.
At zero magnetic field the Eq.(\ref{pi00}) defines the Debye screening
radius, $r_D^{-2} =\Pi_{00}(H=0,p_0=0,{\bf p}\rightarrow \nolinebreak 0)$
\cite{frad}, but this is not valid for $\mu$ and $H\ne0$.
\par
The other two components $\Pi_{01}$ and $\Pi_{02}$ and their
conjugates may be evaluated by taking the derivatives of the
fermion density with respect to magnetic field in the static limit
\cite{zeit,frad,roja}, which has been discussed more explicitly
in Ref.\cite{zeit}:

        \begin{equation}
        \Pi_{0j}(p \rightarrow 0) = i e\/ \varepsilon_{ij}p_i\frac{\partial
        \rho}{\partial H} \quad i,j = 1,2~.
        \end{equation}

\par
It has been shown in Ref.\cite{zeit} that the components $\Pi_{0j}$  describe a
conductivity in the plane orthogonal to the magnetic field  which is
Hall-like \cite{zeit,zeit1}:

        \begin{equation}
        \sigma_{ij} =
        = i \left. \frac{\partial \Pi_{0i}(p)}{\partial p_j}
                \right|_{p \rightarrow 0}
        = e \varepsilon_{ij} \frac{\partial \rho}{\partial H}
        \quad i,j = 1,2.
        \label{sigma}
        \end{equation}

\par
Using Eq.(\ref{sigma}) and the expression for the fermion density
Eq.(\ref{rho}), one has

        \begin{eqnarray}
        \lefteqn{\Pi_{0j}(p_0, {\bf p} \rightarrow 0) =
        \frac{i \varepsilon_{ij}p_i}{\pi^2}
        \sum_{n=0}^{\left[  \frac{\mu^2 - m^2}{2eH} \right]}b_n
        \times}\nonumber\\
        &&
\left\{
        \frac{\mu^2-m^2-3eHn}{(\mu^2-m^2-2eHn)^{1/2}} -
        \frac{T^2\pi^2}{12}
\left(
        \frac{(\mu^2-m^2-2eHn)(m^2 + 2eHn) + eHn
(2\mu^2+m^2+2eHn)}{(\mu^2 - m^2- 2eHn)^{5/2}}
\right)
\right.\nonumber\\
& &\nonumber\\
&& \left.
+ \frac{\alpha}{2\pi} \mu
\left[
\left(
\ln \frac{\mu^2 -m^2 -2eHn}{eH}
- \frac{2eHn}{\mu^2 -m^2 -2eHn}
\right)
\right.\right.\nonumber\\
& &\nonumber\\
&& \left.\left.
- \frac{T^2\pi^2}{6}
\left(
\frac{2eHn (\mu^2+3m^2+6eHn) +
(\mu^2 -m^2 -2eHn) (3m^2 +6eHn -\mu^2)} {(\mu^2 -m^2 -2eHn)^3}
\right)
\right]
   \right\}.
        \label{hall}
        \end{eqnarray}
\par
It has been pointed out in Ref.\cite{zeit} that the above expression is
the Hall conductivity in the
$QED_{3+1}$ is an oscillating function of the chemical potential and the
magnetic field in one -loop corrections, which are close to
``giant oscillations'', well-known in condensed matter physics\cite{abri}
and resonant effects in $QED$\cite{shab} and semiconductors\cite{koro}.
Thus, with inclusion of two -loop corrections to one -loop corrections, we
found from figure 3 that it enhances the oscillations further.
\section{Weak magnetic field limit}
In the weak magnetic field limit, the state $n$ can be replaced in
the limit $2eHn = \theta$
\begin{equation}
\sum_{n=0}^{\left[ \frac{\mu^2 -m^2}{2eH} \right]}
= \int_{0}^{\left[ \frac{\mu^2 -m^2}{2eH} \right]} d\!~~ n
= \lim_{H\rightarrow 0} \frac{1}{2eH} {\int_{0}^{(\mu^2-m^2)}} d
\!~~ \theta
\label{H0}
\end{equation}
Substituting Eq. (\ref{H0}) in Eq. (\ref{eqT0}), we have the effective
Lagrangian as:

        \begin{eqnarray}
        \lefteqn{\tilde{\cal L}^{\rm eff}(T=0, H_0 \ll (\mu^2 -m^2), \mu)
} \nonumber\\
       && \approx \frac{1}{(2\pi)^2}
   \left[
\frac{\mu (2\mu^2 - 5m^2) ( \sqrt{\mu^2 -m^2})}{6}
    + \frac{m^4}{2} \ln \frac{\mu+\sqrt{\mu^2-m^2}}{m}
\right.\nonumber\\
&& \left.
+ \frac{\alpha}{2\pi}
\frac{(\mu^2 -m^2)^2}{2}
\left[
\ln \|\frac{m^2 -\mu^2}{eH_0}\|  -\frac{1} {2}
\right]
\right]
        \end{eqnarray}
The fermion density is calculated from Eq. (\ref{rho}) for weak
field limit at zero temperature by substituting $2eHn = \theta$, we get

 \begin{eqnarray}
        \lefteqn{\rho(H_0\ll(\mu^2-m^2),\mu,T=0) }
\nonumber\\
&&  \approx \frac{1}{2\pi^2}
\left\{
\frac{2}{3}(\mu^2-m^2)^{3/2}
\right.\nonumber\\
&& \left.
+\frac{\alpha}{2\pi}
\left[
\mu (\mu^2 -m^2)\ln\| \frac{\mu^2 -m^2}{eH_0}\|
\right]
\right\}.
\label{rho0}
       \end{eqnarray}
\par
In figure 5, we plot the ratio fermion density $\rho(H,\mu,T=0)/\rho(H_0
\ll(\mu^2-m^2),\mu,T=0)$ as a function of chemical potential with fixed
magnetic field, (a) $H\approx 5 \times 10^{15}$ G and (b) $H\approx 10^{16}$ G.
We choose the weak field limit $H_0$ to be $4.4\times 10^{13}$ G.
As we increase the chemical potential, the ratio
decreases and reaches  to unity for both curve (a) and (b). We noticed that
the magnetic field has significant contribution to the fermion density and
hence to the total effective Lagrangian.
\section{Conclusion}
In conclusion, we discussed the effect of the two - loop $QED$
effective Lagrangian at finite temperature and density in a
constant external magnetic fields. In particularly following
Ref.\cite{zeit}, we compute
the low - temperature corrections to the effective Lagrangian.
Also, we improved the calculation by including the two - loop
corrections in the fermion density, magnetization and the some
components of the polarization operators responsible for the
Hall conductivity and Debye screening in a finite fermion
density $QED_{3+1}$ with uniform magnetic fields. At the end, we
derived the effective Lagrangian in the weak field limit at zero
temperature. In this paper, we found that the two - loop
corrections has a significant contribution to the fermion
density and magnetization in variation with magnetic fields for
higher chemical potentials. So, we do expect to the de Hass -
van Alphen oscillations in the astrophysical objects like
neutron stars and will be published elsewhere. Finally, it
would be some interest to extend this work in $QED_{2+1}$. Also,
one could treat the present calculation in slowly varying electric
and magnetic fields.
\section{Acknowledgement}
I would like to thank R. Parwani for introducing this problem. It
is great pleasure to thank S. Mohanty for useful
discussions and  Per Elmfors for comments.

\vfil
\eject
\newpage
\begin{figure}
\caption { The ratio of total effective Lagrangian ${\cal L}_{tot}$
(including two
 - loop corrections) to one - loop effective Lagrangian ${\cal L}_{1}$ as a
function of magnetic field for fixed chemical potential. Where curves
(a), (b) and (c) are  for $\mu=$ 2, 5 and 10 MeV respectively.}
\end{figure}
\begin{figure}
\caption { The fermion density as a function of the
chemical potential for fixed magnetic field ($H\approx 10^{15}$
G) with (a) one - loop corrections and (b) one -loop and two
-loop corrections.}
\end{figure}
\begin{figure}
\caption { The fermion density vs magnetic field
for fixed chemical potential ($\mu=10$ MeV). Curves (a) and
(b) are same as figure 2.}
\end{figure}
\begin{figure}
\caption { The magnetization as a function of magnetic field
for fixed chemical potential ($\mu =10$ MeV). Curves (a) and
(b) are same as figure 2.}
\end{figure}
\begin{figure}
\caption {The ratio fermion density $\rho(H,\mu,T=0)/\rho(H_0
\ll(\mu^2-m^2),\mu,T=0)$ as a function of chemical potential with fixed
magnetic field, (a) $H\approx 5\times 10^{15}$ G and (b) $H\approx 10^{16}$ G.
Here the weak field limit $H_0$ to be $4.4\times 10^{13}$ G.}
\end{figure}
\vfil
\eject
\end{document}